\documentclass[preprint,review,12pt]{elsarticle}

\usepackage{amssymb}
\usepackage{amsmath}
\usepackage{graphicx}
\usepackage[retainorgcmds]{IEEEtrantools}
\usepackage[colorlinks]{hyperref}
\hypersetup{
  citecolor = {blue}
}
\usepackage{siunitx}
\usepackage{pdflscape}

\usepackage{enumitem}
\usepackage[normalem]{ulem}
\useunder{\uline}{\ul}{}
\usepackage{setspace} 
\usepackage[utf8]{inputenc}
\usepackage[linesnumbered, ruled, vlined]{algorithm2e}
\usepackage{rotating}
\usepackage{nicefrac}
\usepackage{xspace}
\usepackage{float}
\usepackage{caption}
\usepackage{subcaption}
\usepackage{rotating, lscape, longtable, tabu, booktabs, siunitx, multirow}
\usepackage[capitalise, noabbrev]{cleveref}
\usepackage{verbatim}
\usepackage{mathtools}

\usepackage{tkz-graph}

\usetikzlibrary{arrows,arrows.meta,shapes,decorations.pathmorphing, decorations.markings, positioning}
\let\svtikzpicture\tikzpicture
\def\tikzpicture{\noindent\svtikzpicture}

\newcommand{\uset}[1]{\ifmmode\left\{\,#1\,\right\}\else\{\,#1\,\}\fi}
\newcommand{\ulst}[1]{\ifmmode\left[\,#1\,\right]\else[\,#1\,]\fi}
\newcommand{\upar}[1]{\ifmmode\left(\,#1\,\right)\else(\,#1\,)\fi}
\newcommand{\uioc}[1]{\ifmmode\left(\,#1\,\right]\else(\,#1\,]\fi}
\newcommand{\uico}[1]{\ifmmode\left[\,#1\,\right)\else[\,#1\,)\fi}

\journal{xxxx}

\begin{document}

\begin{frontmatter}

\title{A Survey on Federated Learning Poisoning Attacks and Defenses}

\author[add1]{Junchuan Liang}
\ead{liangjunchuan@stu.sicnu.edu.cn}

\author[add1,add2]{Rong Wang\corref{cor1}}
\ead{rwang@sicnu.edu.cn}

\author[add1]{Chaosheng Feng}
\ead{csfeng@126.com}

\author[add3]{Chin-Chen Chang}
\ead{alan3c@gmail.com}

\address[add1]{School of Computer Science, Sichuan Normal University, Chengdu , China}
\address[add2]{Visual Computing and Virtual Reality Key Laboratory of Sichuan, Sichuan Normal University, Chengdu, China}
\address[add3]{Department of Information Engineering and Computer Science, Feng Chia University, Taichung 40724, China}

\cortext[cor1]{Corresponding author}


\begin{abstract}
As one kind of distributed machine learning technique, federated learning enables multiple clients to build a model across decentralized data collaboratively without explicitly aggregating the data. Due to its ability to break data silos, federated learning has received increasing attention in many fields, including finance, healthcare, and education. However, the invisibility of clients’ training data and the local training process result in some security issues. Recently, many works have been proposed to research the security attacks and defenses in federated learning, but there has been no special survey on poisoning attacks on federated learning and the corresponding defenses. In this paper, we investigate the most advanced schemes of federated learning poisoning attacks and defenses and point out the future directions in these areas.
\end{abstract}


\end{frontmatter}


\section{Introduction}
\label{sec:intro}
With the development of global digitalization, a huge amount of data is being generated every day, and techniques of machine learning have emerged to identify the valuable information in this data. A large amount of data may be distributed across different devices and may contain significant amounts of private data, and this makes centralized machine learning difficult. To overcome the problem, federated learning \cite{1,2,3} has been proposed to build machine learning models on decentralized data in a collaborative way. Its emergence provides a promising solution to solve the data silos, without compromising local data privacy. Since federated learning was proposed in 2016 \cite{102}, it has been applied to many applications, such as natural language processing (NLP) \cite{4,5,6,7}, healthcare \cite{8,9,10,11,12,13,14,64}, and the Internet of Things (IoT) \cite{15,16,17}.

\begin{figure*}[h]
	\centering
	\includegraphics[width=1\textwidth]{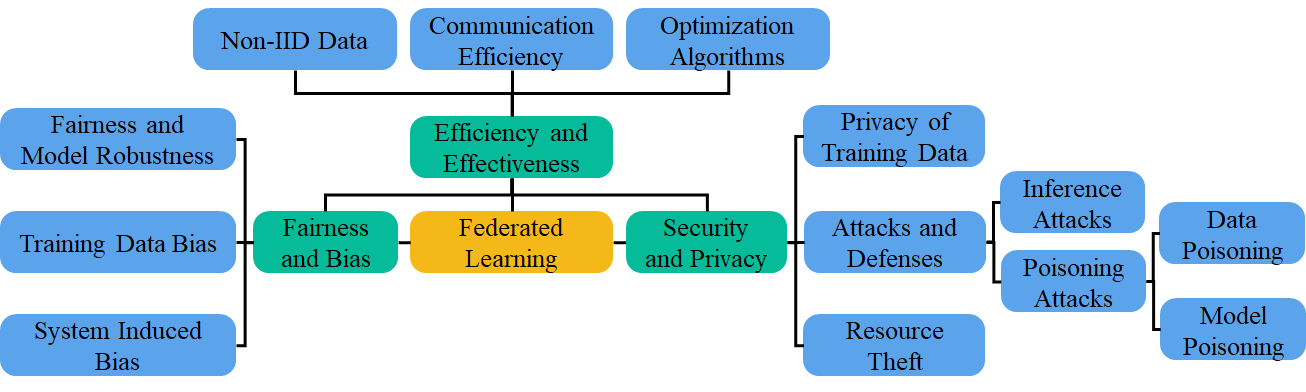}
	\caption{Main research directions of federated learning}
	\label{fig1}
	\end{figure*}

As shown in Fig. \ref{fig1}, research on federated learning can be categorized into three groups, i.e., fairness and bias, efficiency and effectiveness, and security and privacy. Specifically, the research on fairness and bias focuses on how to ensure the fairness of the model in the case of inconsistent data distribution among clients; the research on efficiency and effectiveness has been focused on determining how to reduce the number of server-client communications as much as possible while preserving the accuracy of the learned model, and the research on privacy and security is concerned about how to ensure the privacy of clients’ data and model security in unsafe environments (e.g., some clients are malicious, or the server is honest but curious). In this paper, we focus on federated learning poisoning attacks and defenses, which belong to the aspect of security and privacy. 

\begin{figure*}[h]%
	\centering
	\includegraphics[width=0.4\textwidth]{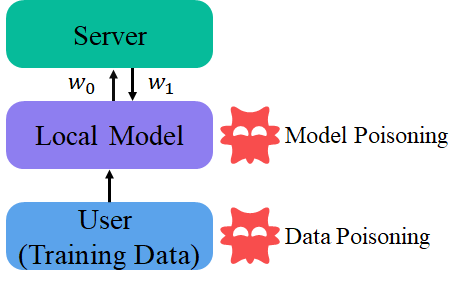}
	\caption{Data poisoning and model poisoning}
	\label{fig2}
\end{figure*}

There are two main types of attacks in federated learning, i.e., the inference attack and the poisoning attack. The goal of the inference attack is to infer or recover local data of other clients through the parameters of the global model. The attacker in an inference attack may be a malicious client, an honest but curious server, or an eavesdropper in a communication channel. The goal of the poisoning attacker is to corrupt the performance of the model or to exploit the federated learning framework for illegal purposes. In poisoning attacks, the attacker is located mainly on the client side. Because the clients’ local data and training process are invisible to each other in federated learning, the poisoning attack has plenty of room for survival and stealth. As shown in Fig. \ref{fig2}, the poisoning attack can be divided into two types, i.e., data poisoning and model poisoning. Data poisoning refers to the malicious client in federated learning who either actively or passively uses some poisoned data for the training model, and model poisoning means that an attacker tampers the parameters of his/her local model, which further affects the parameters of the global model. The poisoning attacker is free to modify the model’s parameters according to her/his needs for poisoning. Recent studies have demonstrated that model poisoning poses a more serious threat to the security of federated learning than data poisoning \cite{18,19}. The intuition of this viewpoint is that the final result of data poisoning will lead to changes in the parameters of the model. Therefore, in this paper, poisoning attacks refer to both data poisoning and model poisoning.
	
Since the introduction of federated learning, there has been an increasing amount of research work on security and privacy in federated learning. The descriptions of some recent surveys on this topic are provided as follows.
	
(1) Peter Kairouz et al. \cite{20} provided a detailed introduction to the concepts, application scenarios, and challenges of federated learning. These challenges include improving efficiency and effectiveness, preserving data privacy, defending against attacks, keeping fairness, and developing software and datasets for federated learning.
	
(2) Mothukuri et al. \cite{21} introduced the problems of communication, poisoning attacks, inference attacks, and privacy leakage in federated learning.
	
(3) Yilun Jin et al. \cite{22} introduced the application of semi-supervised federated learning and summarized the implementation of semi-supervised federated learning.
	
(4) Alberto Blanco-Justicia et al. \cite{23} reviewed the federated learning poisoning models, privacy inference, and the methods of defenses.
	
(5) Wang et al. \cite{24} focused on the detailed classification of privacy-preserving federated learning schemes and the corresponding defenses.
	
(6) Zhilin Wang et al. \cite{25} presented the implementation methods of blockchain-based federated learning and their advantages and disadvantages in applications.
	
(7) Mengna Yang et al. \cite{26} conducted a survey on privacy protection methods, communication overhead solutions, and malicious participant defense methods in federated learning.
	
(8) Zhilin Wang et al. \cite{27} summarized the defense methods regarding model poisoning attacks.
	
(9) Lingjuan Lyu et al. \cite{28} summarized recent federated learning privacy-preserving aggregation schemes.
	
(10) Ziyao Liu et al. \cite{29} summarized federated learning privacy attacks and defense schemes, and they also presented the current challenges related to this area.
	
(11) Nuria Rodríguez-Barroso et al. \cite{30} focused on the robustness, privacy attacks, and defenses of federated learning.
	
Although the surveys above summarize some privacy attacks and defenses in the literature concerned with federated learning, in the last five years there has been no specialized review on poisoning attacks or the corresponding defense mechanisms. In this paper, we combine the state-of-the-art works to make a comprehensive survey on federated learning poisoning attacks and defenses, and we point out the future directions of this aspect.

\begin{figure*}[h]%
	\centering
	\includegraphics[width=1\textwidth]{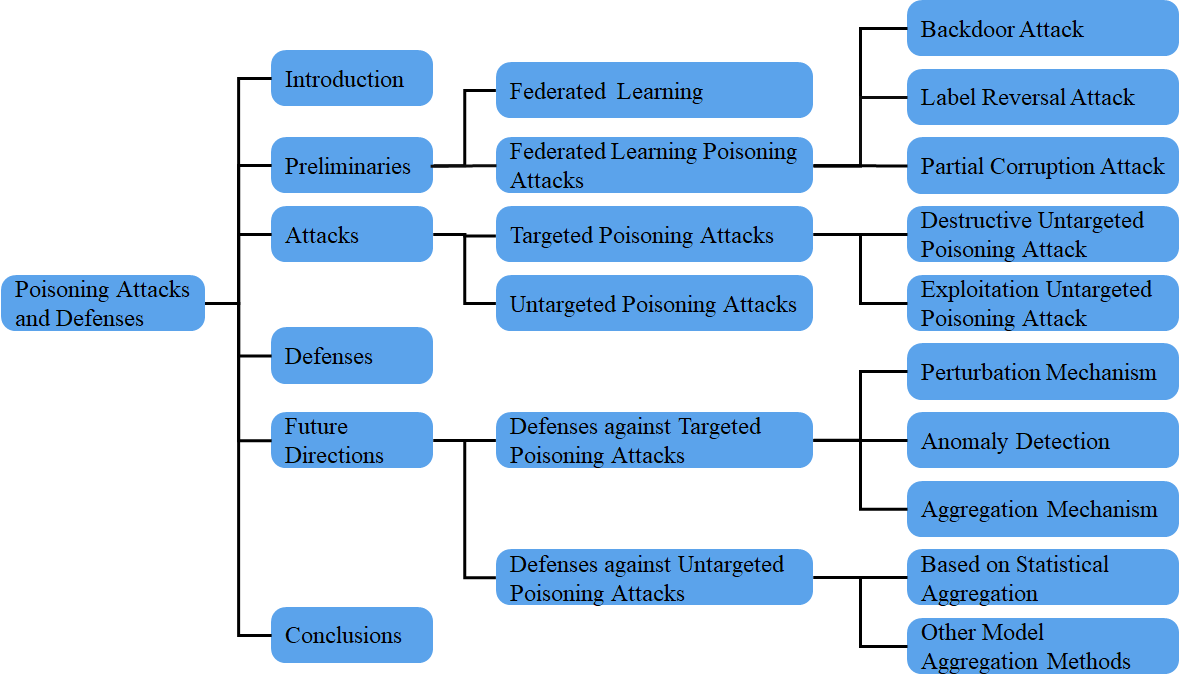}
	\caption{Organization of this paper}
	\label{fig3}
\end{figure*}

This paper is organized as shown in Fig. \ref{fig3}.
	
\section{Preliminaries}\label{sec2}
\subsection{Federated learning}\label{subsec2.1}
Federated learning is a distributed machine learning approach where a central server collaborates with multiple clients, such as mobile devices, to train a global model without exchanging raw data.
	
Let's denote the global model parameters by $\theta$, and let $U_1, U_2, \dots, U_n$ be the $n$ clients participating in the federated learning process. Each client $U_i$ has its own local dataset $D_i$ and trains a local model with parameters denoted by $\theta_i$. The local models are then sent to the central server, which aggregates them to obtain an updated global model, as follows:
	
\begin{equation}
	\theta_{t+1} = \sum_{i=1}^n \frac{|\mathcal{D}i|}{|\mathcal{D}|} \theta_{i,t}
\end{equation}
where $\theta_{i,t}$ is the local model of client $U_i$ at iteration $t$, $|\mathcal{D}_i|$ is the size of the local dataset of client $U_i$, and $|\mathcal{D}|$ is the total size of all local datasets.
	
The aggregation process can be performed using various methods, such as a simple averaging or weighted averaging based on the local dataset sizes. The updated global model $\theta_{t+1}$ is then sent back to the clients, and the process is repeated until convergence.
	
Federated learning can be formulated as the following optimization problem:
\begin{equation}
	\min_{\theta} \sum_{i=1}^n \frac{|\mathcal{D}_i|}{|\mathcal{D}|} \mathcal{L}_i(\theta)
\end{equation}
where $\mathcal{L}_i(\theta)$ is the loss function of client $U_i$ with respect to the global model parameters $\theta$. The goal of federated learning is to find the optimal global model parameters $\theta$ that minimize the average loss over all clients.
	
\begin{figure*}[h]%
	\centering
	\includegraphics[width=0.6\textwidth]{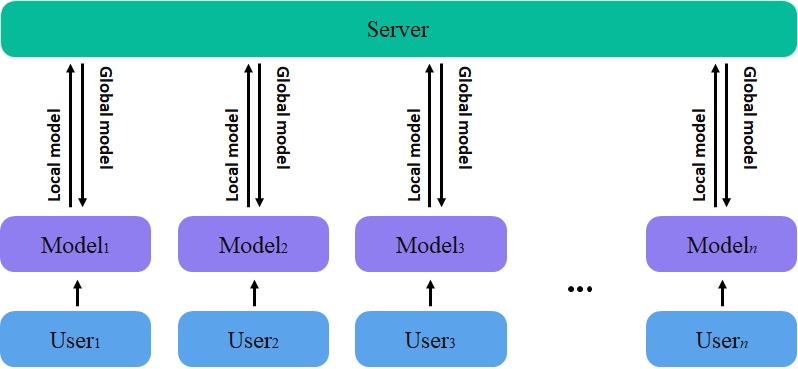}
	\caption{Framework of typical federated learning}
	\label{fig4}
\end{figure*}

As shown in Fig. \ref{fig4}, in the typical federated learning the server first initializes the parameters of the global model and distributes them to the clients who are participating in federated learning; then the clients train their local models using their own data, and they upload the new parameters of their models to the server for aggregation. The above process is repeated until the pre-defined condition is met.
		
\begin{figure*}[h]%
	\centering
	\includegraphics[width=0.8\textwidth]{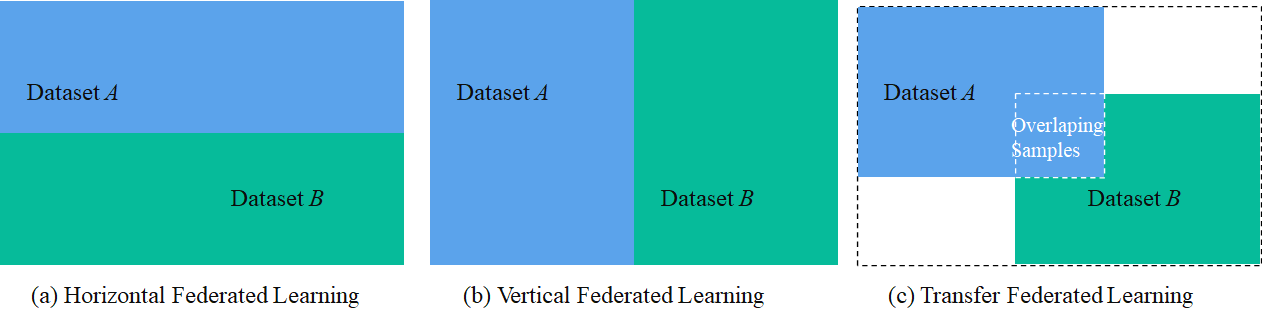}
	\caption{Categories of federated learning}
	\label{fig5}
\end{figure*}

According to the distribution of training data, federated learning can be categorized into the three groups shown in Fig. \ref{fig5}, i.e., horizontal federated learning \cite{32}, vertical federated learning \cite{33,65}, and transfer federated learning \cite{34}.
	
\begin{itemize}
	\item[$\bullet$] Horizontal federated learning is applicable to the cases in which different data sets share the same feature space but differ in sample ID.
	\item[$\bullet$] Vertical federated learning is applicable to the cases in which different data sets share the same sample ID but differ in feature space.
	\item[$\bullet$] Transfer federated learning is applicable to the cases in which different data sets do not share the same sample ID, nor do they have the same features.
\end{itemize}

\subsection{Poisoning attacks on federated learning }
\label{subsec2.2}
\begin{figure*}[h]%
	\centering
	\includegraphics[width=1\textwidth]{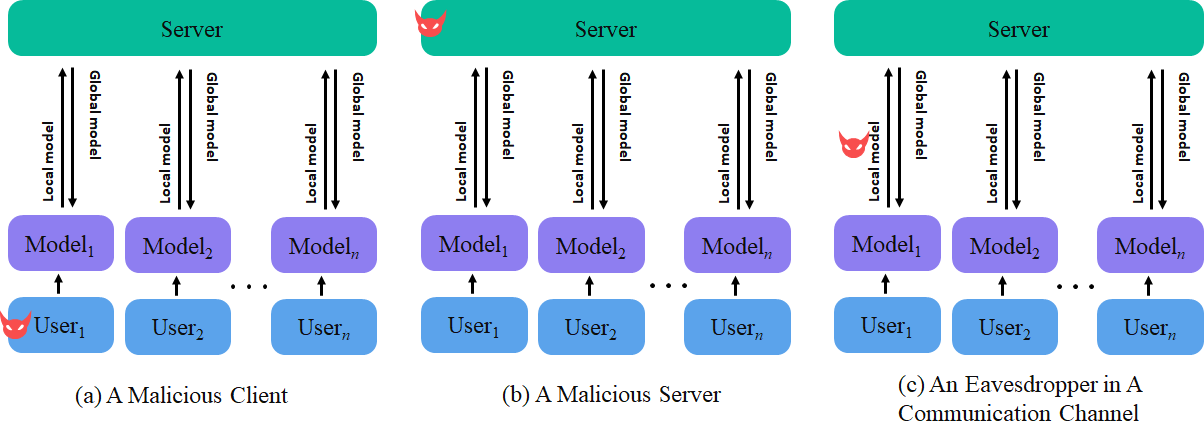}
	\caption{Possible locations of poisoning attackers}
	\label{fig6}
\end{figure*}

The term ``poisoning attack'' refers to a case in which an attacker achieves illegal purposes by tampering with the training data or the parameters of the model. The goal of the poisoning attack is to destroy model aggregation or achieve illegal purposes (such as stealing models) through the collaborative training process. Namely, it destroys the integrity and availability of the learning system. By contrast, the inference attack destroys the secrecy of the learning system. In terms of the locations of an attacker, as shown in Fig. \ref{fig6}, the attacker can be on the client-side, the server-side, or in the communication channel. Among the three potential locations of the attacker, the most common case is the poisoning attack occurring on the client-side.
	
Poisoning attacks commonly occur in the field of machine learning. The attacks arise because of the uninterpretability of machine learning. Federated learning is one kind of distributed machine learning technique, so it also could incur the threat of poisoning attacks. In addition, compared with the centralized machine learning, federated learning is at greater risk of poisoning attacks because the client-side training data and the local training process are invisible to each participant.
		
\section{Poisoning attacks}
\label{sec3}
\begin{figure*}[h]%
	\centering
	\includegraphics[width=0.8\textwidth]{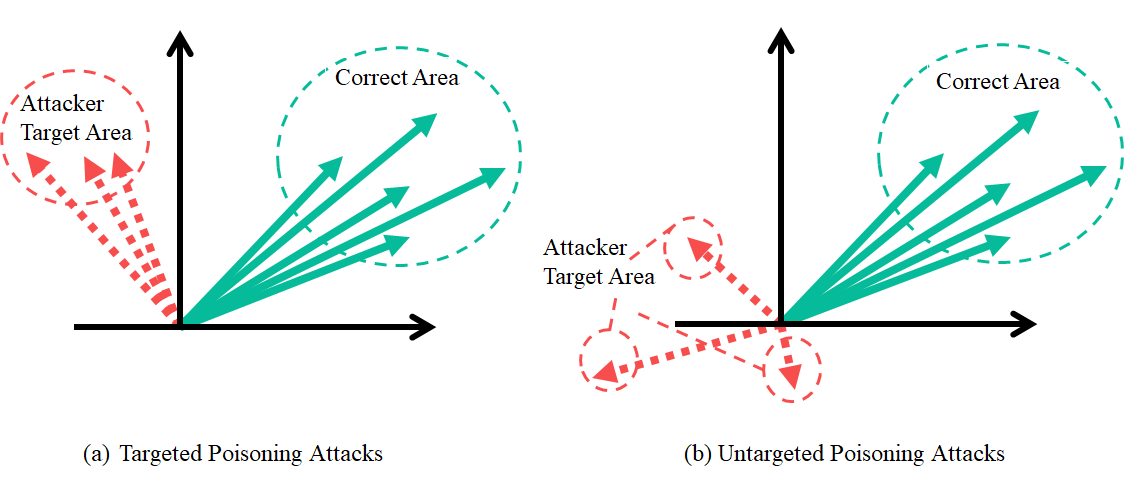}
	\caption{Targeted and untargeted poisoning attacks}
	\label{fig7}
\end{figure*}

Poisoning attacks can be divided into two types, i.e., targeted poisoning attacks and untargeted poisoning attacks. Figure. \ref{fig7} shows the difference between the two types. In this figure, the green circle represents the area where a benign model is expressed, and the green arrows in the circle represent the directions of data vectors updated by normal clients. By contrast, the red circle represents the area where the poisoned model is expressed, and the red arrows represent the directions of data vectors updated by attackers. In other words, the case that the attacker conducts an attack with a specified purpose can make the red arrows in Fig. 7(a) share the same direction; in contrast, the red arrows in Fig. 7(b) have different directions when the attacker conducts an untargeted attack. Table \ref{table2} presents the classification and threat level of poisoning attacks.
	
\begin{center}
	\begin{table*}[ht]  
		\centering 
		\caption{Description of poisoning attacks}
		\label{table2}
		\begin{tabular}{p{2cm}p{2.2cm}p{1.6cm}p{6cm}}
			\toprule   
			Type of Attack & Method & Risk level&Description of the attack \\  
			\midrule   
			Target poisoning attack & Backdoor attack & High & backdoor triggers are embedded in the parameters of models that are uploaded to the server \cite{35,36,37,38,39,40,41,42,43,44,49,71,72,73,74}\\  
			\\
			& Label reversal attack& High & to control clients to predict a sample with a pre-defined label \cite{51,52,53,32,55}\\   
			\\
			& Partial corruption attack& Medium & a task of the model is corrupted \cite{38,53}\\
			\\
			\toprule 
			Untargeted poisoning attack & Disruption untargeted poisoning attack & High & to make it impossible for the model to converge \cite{19,75,76,77}\\
			\\
			& Utilization untargeted poisoning attack & Low & to use the global models for Secret Communication \cite{78,79}\\
			\bottomrule  
		\end{tabular}
	\end{table*}
\end{center}

\subsection{Targeted poisoning attacks}
\label{subsec3.1}
The aim of a targeted poisoning attack is to completely control or affect one subtask of the federated learning model without affecting any other subtask. For example, in the handwritten digit recognition task, the recognition of any handwritten digit can be treated as a subtask. Typically targeted poisoning attacks include the backdoor attack, the label reversal attack, and the partial corruption attack.
	
\subsubsection{Backdoor attacks}\label{subsubsec3.1.1}
The backdoor attack aims to affect the performance of a model on a particular subtask while maintaining the original performance on other subtasks. The backdoor attack injects triggers into the clients’ training data to poison the model. Since this attack does not change the label of the training data, the attack is known as the ``clean label attack'' in \cite{103}. According to the type of triggers, the backdoor attack can be divided into two groups, i.e., the numerical backdoor \cite{35,36,37,38,39,40} attack and the physical backdoor attack \cite{41,42,43,44}. The focus of the numerical backdoor attack is on tampering with certain samples in the training dataset such that the tampered samples are learned by the federated model as normal samples. Physical backdoor attacks are those in which the backdoor trigger appears to be real. A model that has been subjected to a backdoor attack does not behave abnormally under normal circumstances. However, once the backdoor is triggered by a certain test sample, the prediction result of the test sample can be controlled by the attacker.
		
Bhagoji et al. \cite{37} proposed a targeted poisoning attack on federated learning conducted by a single malicious client. The goal of their attack is to make the model misclassify the target label with high confidence, without affecting the classification of other labels. However, their method of boosting the learning rate in this paper can lead to the problem of catastrophic forgetting, and this attack requires the attacker to participate in each round of federated learning in order to maintain the backdoor accuracy. Sun et al. \cite{39} and Bagdasaryan et al. \cite{36} focused on model poisoning attacks based on the assumption that the attacker has a large amount of knowledge. The attacker tries to replace an entire model with the backdoor model by observing the updated parameters in the $t$-th round of training. Wang et al. \cite{40} analyzed backdoor attacks and the problem of robustness in federated learning, and they proposed a new backdoor attack, which they called an edge-case backdoor. The edge-case backdoor attack forces the model to misclassify seemingly easy inputs. These inputs are present in the tails of the input distribution. Bagdasaryan et al. \cite{35} proposed a backdoor attack that does not control the classification output but changes the final output of some specific task. For example, an attacker can control the model to output a prediction of label ``3'' even when the input image corresponds to label ``1'' or label ``2''. Jagielski et al. \cite{38} proposed a subgroup backdoor attack. This attack has an advantage over existing backdoor poisoning attacks because it does not require any modification of the test data to induce misclassification. Xie et al. \cite{48} proposed a distributed backdoor poisoning attack, DBA, in which a particular backdoor trigger is dispersed into federated learning clients. It was found in this paper that the distributed poisoning approach with global triggers has a higher success rate than that of a single backdoor because the former uses synthetic triggers to manipulate the output of the model.
	
Eykholt et al. \cite{42} and Du et al. \cite{41} proposed a physical backdoor attack based on a convolutional neural network (CNN). This attack mainly acts on the downstream tasks of machine vision, including target detection and target segmentation, and it can result in the misclassification of test samples. Xin et al. developed the DPATCH \cite{43} method, which is based on a physical backdoor attack that compromises the target detection. DPATCH uses the generation of noisy backdoors on the test data and optimizes the backdoors through a loss function to enhance the performance of the attack. Without achieving any real training data, Zhang et al. \cite{44} generated new training data that were similar to the real training data by using a generative adversarial network (GAN), and they successfully conducted the backdoor attack \cite{71}. Similarly, Zhang et al. \cite{49} used GAN to conduct a stealthier poisoning attack.
	
The key that contributes to the success of backdoor attacks is the enormous capacity of modern deep learning models. The attacker takes advantage of this capability to inject covert backdoors without significant impact on the accuracy of the model. Research on backdoor attacks will contribute to the development of interpretability in the field of deep learning.

\subsubsection{Label reversal attacks}
\label{subsubsec3.1.2}
The label reversal attack \cite{50,51,52,53,48} is known as the sibyl attack \cite{50} or dirty label attack. A label reversal attack refers to a case in which the attacker controls the labels of the training data owned by clients. Unlike the backdoor attack, which poisons the training data, a label reversal attack hijacks the labels of the training data in a particular classification task. The attacker only tampers the labels of samples and does not change their feature values. For example, in the task of handwritten digit recognition, some of the clients holding a dataset that contains samples labeled as ``1'' are hijacked, and the recognition result of these samples is modified to label ``7''. This causes the model to incorrectly treat the number ``1'' as the number ``7'' during training. It is a strong assumption that most of the data have been hijacked. The label reversal attack has been proposed as a successful attack in the field of machine learning \cite{55,32}. Fung et al. \cite{50} and Tolpegin et al. \cite{53} investigated the application of label reversal attacks in federated learning scenarios. Tolpegin et al. \cite{53} discussed the implementation of a label reversal attack when a small number of clients (10$\%$ - 40$\%$) are hijacked. Due to its simple assumption and good attack performance, the label reversal attack is one of the important targeted poisoning attacks in federated learning. The difficulty of this research is how to control the number of clients as little as possible to achieve the same attack performance.
	
\subsubsection{Partial corruption attacks}
\label{subsubsec3.1.3}
The partial corruption attack is designed to make a particular type of subtask less successful without affecting the other tasks. The attack is similar to but not as strong as the backdoor attack. In general, it is believed that the control of a machine learning model is more difficult than indiscriminate re-bye corruption of the model. Tolpegin et al. \cite{53} proposed this method of attack and experimentally demonstrated that hijacking a small number of clients could cause a significant decrease in the accuracy of a subtask.
	
The subgroup poisoning attack proposed by Jagielski et al. \cite{38} implements both backdoor and partial corruption attacks. The partial corruption attack in federated learning is just beginning to be explored, and there are many worthwhile directions for future research. For example, attackers should consider how to increase the number of corruptions or circumvent state-of-the-art defense mechanisms.
	
\subsection{Untargeted poisoning attacks}
\label{subsec3.2}
The purpose of untargeted poisoning attacks is to corrupt or manipulate a federated learning model by changing the training data or the parameters of the model. This allows the attacker to control some parameters of the training process of some selected clients, such as raw data, training rounds, and model parameters.
	
\subsubsection{Disruption untargeted poisoning attacks}
\label{subsubsec3.2.1}
The main goal of this attack is to corrupt the federated learning model so that the training process of the model does not converge successfully. Carlini \cite{75} pointed out that the attacker can use only 0.1$\%$ of the training data to conduct this attack. Federated learning can filter a certain degree of anomalous data due to its aggregation rules (e.g., federated averaging \cite{105}). Baruch et al. \cite{76} crafted an untargeted attack based on a reverse aggregation vector to make the federated learning model aggregation collapse. This attack renders some statistical-based defenses ineffective. Experiments \cite{76} have shown that 15$\%$ of malicious clients are sufficient to corrupt the entire model on the MNIST dataset. Xie et al. \cite{77} proposed an inner product manipulation attack. Their core intuition is that the direction of the aggregation vector needs to be the same as the direction of the true gradient to make the gradient descent algorithm effective during training. That is, the inner product of the aggregation vector and the true gradient must be non-negative. Thus, a successful attack can be achieved by manipulating the inner product of the aggregation vector and the true gradient. Fang et al. \cite{19} considered the non-target poisoning attack as an optimization problem. They demonstrated that global models based on Byzantine robust aggregation rules will suffer worse test error rates when the attacker directly manipulates the parameters of the client-side local model. Similarly, by manipulating the client’s data, the attacker in \cite{19} updates its malicious model by maximally perturbing the benign aggregate in the malicious direction, while also evading detection by the robust aggregate algorithm.
	
\subsubsection{Exploitation untargeted poisoning attacks}
\label{subsubsec3.2.2}
An exploitation untargeted poisoning attack refers to a case in which an attacker uses a poisoning attack to exploit the federated learning framework for illegal purposes. Lin et al. \cite{78} proposed a hitchhiking attack. In this attack, the attacker disguises himself/herself as a benign client and gets the global model freely, rather than tampering with the output of the model. This attack does not make a serious impact on the utility of the model, but the speed of the model’s convergence will be affected seriously by this attack. Currently, there are no specific defenses against this type of attack due to its covert purpose and less destructive nature. Costa et al. \cite{79} successfully implemented a 1-bit secret communication between two clients through the poisoning attack. This attack does not significantly affect the aggregation model, but it can be exploited for covert information transfer in this way.

\section{Defenses against poisoning attacks}
\label{sec4}
\subsection{Defenses against targeted poisoning attacks}
\label{subsec4.1}
\subsubsection{Perturbation mechanisms}\label{subsubsec4.1.1}
The perturbation mechanism is a method that perturbs the training data or the model parameters to avoid poisoning attacks on the model. Perturbation mechanisms gain a security improvement in the model at the expense of a small amount of model accuracy. One of the most well-known perturbation mechanisms is differential privacy \cite{80}, which has been used extensively in federated learning \cite{60,81}. The rigorous mathematical theoretical basis and good performance of differential privacy have made it relevant to many studies in security.
	
For any two neighboring datasets that differ in at most one element, the output distributions of a differentially-private algorithm over these two datasets are ``almost the same''~\cite{dwork2014the}. It ensures that an individual's record cannot be identified from the output. The level of privacy is controlled by a parameter $\epsilon$, called the privacy budget. The smaller the value of $\epsilon$, the stronger the privacy guarantee. A randomized algorithm $\mathcal{A}$ satisfies $(\epsilon, \delta)$-differential privacy if for any two neighbouring datasets, $D$ and $D'$, and for any subset of outputs $S \subseteq Range(\mathcal{A})$, it satisfies:
	
\begin{equation}
	\label{eq3}
	\Pr[\mathcal{A}(D) \in S] \leq e^{\epsilon} \Pr[\mathcal{A}(D') \in S] + \delta
\end{equation}
where $\epsilon$ controls the privacy level and $\delta$ allows a small probability of privacy loss. The privacy budget $\epsilon$ is consumed when the differentially-private algorithm is running, and the total privacy loss is the sum of the values of $\epsilon$ from multiple runs.
	
Du et al. \cite{82} showed that differential privacy can improve the performance of outlier detection. Also, differential privacy can effectively mitigate target poisoning attacks. Doan et al. \cite{83} used perturbed input to detect backdoors in the training data. The idea of their perturbation mechanism is to superimpose the training data. The image backdoor triggers superimposed will become strong features due to position fixation. This technique can be used to verify that some malicious backdoors are not embedded in the training data. Nguyen et al. \cite{84} proposed a novel technique, FLAME, for target attack defenses. This technique uses clustering based on HDBSCAN \cite{85} and pairwise cosine similarity to eliminate outliers in client-side updates. Then, adaptive anomaly cropping is used to eliminate salient values in the model updates. Finally, Gaussian noise is added to the obtained parameters to disrupt possible backdoor triggers. Experiments in \cite{84} have demonstrated that FLAME provides better performance of defense against backdoor attacks compared with the previous methods. However, FLAME requires extensive and complex modifications to the federated learning framework.
	
\subsubsection{Anomaly detection}\label{subsubsec4.1.2}
In the federated learning literature, anomaly detection schemes can be divided into two categories, i.e., clustering-based detection and comparison-based detection. The details of these two types are shown in Table \ref{table3}. The former is mainly unsupervised learning, while the latter is supervised learning and requires a ``clean'' or ``dirty'' parameter as a comparison.
	
\begin{center}
	\begin{table*}[ht]  
		\centering 
		\caption{Defenses against target poisoning attacks}
		\label{table3}
		\begin{tabular}{p{2cm}p{6cm}p{4.2cm}}
			\toprule   
			Defensive Strategies&Defensive Programs&Defensive Targets  \\  
			\midrule   
			Clustering-based Detection& HDBSCAN, cosine similarity \cite{85} & Backdoor attack, label reversal attack, and destructive untargeted attack \\  
			& $k$-Means \cite{86}&Label reversal attack\\    
			&Outlier detection \cite{57,59}&Label reversal attack and backdoor attack \\
			&Spectral clustering \cite{58}&Backdoor attack\\
			&Cosine similarity \cite{50,88}&Label reversal attack\\
			&Kernel density clustering \cite{89}&Label reversal attack\\
			\toprule 
			Comparison-based detection&Comparison of early training samples \cite{54,90} &Backdoor attack\\
			&Vertical self-renewal comparison \cite{91}&Backdoor attack\\
			&Server validation dataset \cite{92} & Backdoor attack, label reversal attack, and destructive untargeted attack\\
			&Cross-sectional comparison and voting \cite{62}&Backdoor attack\\
			\bottomrule  
		\end{tabular}
	\end{table*}
\end{center}
	
\textbf{Clustering-based detection.} The clustering-based anomaly detection scheme is based on the premise that malicious clients have a different distribution of updated parameters because their purpose is not the same as that of benign clients. Shen et al. \cite{86} used $k$-Means to distinguish benign clients from malicious clients. Steinhardt et al. \cite{57} used an outlier detection method to identify malicious clients. Wang et al. \cite{59} used outlier detection to mitigate backdoor attacks in CNNs. Tran et al. \cite{58} and Li et al. \cite{87} used spectral clustering to identify backdoor triggers. Fung et al. \cite{50} considered the parameters of clients’ model updates in federated learning as a vector and used cosine similarity clustering to distinguish malicious clients from benign clients. Khazbak et al. \cite{88} used cosine similarity and went further to protect clients’ data privacy by using differential privacy schemes in the training phase. The experiments \cite{88} demonstrated that it is feasible to perform anomaly detection to distinguish between malicious and benign clients while taking into account the protection of clients’ data privacy. Li et al. \cite{89} used the kernel density clustering \cite{106} to group similar client parameters and then used an outlier detection mechanism to avoid model poisoning.
	
\textbf{Comparison-based detection.} The core of comparison-based detection is to find the parameters of the model trained on ``clean'' or ``dirty'' data. Shen et al. \cite{54} observed that clean samples correspond to a higher correct rate of accuracy in the initial stage of training while samples with backdoor triggers result in a lower correct rate. Therefore, a proposal is made to detect dirty samples by their initial correctness. Based on the observations of \cite{54}, Andreina et al. \cite{90} proposed an early sample verification method to identify clients carrying backdoor triggers with good results. Ozdayi et al. \cite{91} proposed that if a client carries a backdoor task, then the parameters of the model with a backdoor included in the adjacent two uploads will have a higher similarity, compared with the parameters of modes owned by the benign clients. This approach reduces the learning rate of the model of clients that may carry backdoor triggers, thus reducing the impact of backdoor triggers on the global model. Cao et al. \cite{92} further assumed that the server has a small ``clean'' verification dataset. By comparing the verification dataset, their method achieves better-federated learning even when there are a large number (60$\%$) of malicious clients. Wu et al. \cite{62} proposed a method based on a pruning technique. The core intuition of this method is that backdoor nodes will not be triggered under normal circumstances. Instead, nodes with backdoor tasks will be activated when a backdoor trigger is input. Thus, a clean network can be obtained by constantly cutting out nodes that are not active under normal training, and these cuts do not affect the performance of the final model. Similarly, Wu et al. \cite{62} designed a new federated learning aggregation architecture based on this idea. The server asks the client to upload the activity of the neural nodes and prunes the inactive nodes in the network model in order to obtain a clean global model.
	
\subsubsection{Aggregation mechanisms}
\label{subsubsec4.1.3}
Note that aggregation mechanisms originally were designed to defend against both targeted attacks and untargeted attacks. For example, the Byzantine tolerant aggregation mechanism \cite{104} mitigates backdoor attacks to some degree, but it does not completely eliminate the impact of backdoor attacks. The initial idea of the aggregation mechanism is based on the assumption that there are a variable numbers of malicious clients in federated learning who upload useless or even poisoned parameters. Defenders have to ensure that the aggregated model parameters are as useful as possible in such a situation. Hong et al. \cite{93} proposed a gradient shaping scheme to achieve clean aggregated models by pruning the salient gradients.
	
\subsection{Defenses against untargeted poisoning attacks}
\label{subsec4.2}
It is easy to detect most untargeted poisoning attacks by observing the accuracy of the model. Due to the existence of security aggregation, it is difficult for the server to know who is the attacker among a large number of clients. Thus, see Table \ref{table4}, new defenses against untargeted poisoning attacks are needed; that is, defenses are required to protect clients’ data privacy and reduce the bad effect of model parameters updated by malicious clients. The Byzantine tolerance mechanism \cite{104} is proposed to solve this problem. The core of this approach is to guarantee the availability and correctness of the final aggregation result in case a malicious client is unavoidable. But at the same time, Byzantine-tolerant distributed learning assumes, in most cases, that the training data are independent identical distribution (IID) and the loss function is convex. It does not perform well when the loss function of the global model is non-convex \cite{88,94}.
	
\begin{center}
	\begin{table*}[ht]  
		\centering 
		\caption{Untargeted poisoning defenses}
		\label{table4}
		\begin{tabular}{p{1.5cm}p{2.8cm}p{6cm}p{1.8cm}}
			\toprule   
			Strategies & Name & Description & Capability \\  
			\midrule   
			Statistics-based defense& Krum \cite{95}& Euclidean distance square scores&Medium \\  
			& Median \cite{97}& Median as new parameter&Medium \\    
			& Trim-mean \cite{96}& Median with extremes removed &Medium \\
			& Bulyan \cite{98} &Multiple Krums&Medium\\
			\toprule   
			Other defense methods& FLTrust \cite{92}& ReLU cosine similarity, weighted mean& High\\
			& DnC \cite{97}& Random subsampling mean, numerical clipping&High \\    
			& Ensemble \cite{19}& Group Training Voting&low \\
			& GAA \cite{100}&Credit scoring, validation dataset&High\\
			& attestedFL \cite{101}&Horizontal and vertical comparison test&Medium\\
			\bottomrule  
		\end{tabular}
	\end{table*}
\end{center}
	
\subsubsection{Defenses based on statistical aggregation}
\label{subsubsec4.2.1}
The statistical-based approach is one of the classical methods for achieving Byzantine fault tolerance. Krum \cite{95} selected one of the $n$ local model updates based on the distance-squared scoring as the global model update. In the Trim-mean \cite{96} method, the server sets a pruning parameter $k$ and deletes the largest $k$ parameters and the smallest $k$ parameters. Then, the server calculates the average of the remaining parameters as the parameter values in the global model update. In Median \cite{97}, the server sorts the available parameters and selects the median as the global parameter value. Both Krum and Median claimed that their defense mechanisms can protect 50$\%$ of the client nodes. Due to the presence of dimensional catastrophe, the defensive power of these methods decreases with the increasing data dimensionality. Moreover, these methods do not guarantee the convergence of the loss function in federated learning. Mhamdi et al. \cite{98} proposed Bulyan, which essentially is a combination of a variant of Krum and the trimmed mean. Pillutla et al. \cite{99} proposed a method based on the classical geometric median, which has been proven experimentally to be more robust than the traditional aggregation strategies.
	
The statistics-based approach has a simple strategy and is the baseline method for many other attacks and defense validations. However, it still has many shortcomings. First, the method is mainly for untargeted attacks and is less effective against targeted attacks. Second, although it is effective for the case where the client’s data are IID, it is ineffective when the client’s data are non-independently identically distributed (Non-IID). Note that the Non-IID of the data is a common case in reality. Finally, Krum and Bulyan are ineffective because they both require the server to calculate the pairwise distances for local model updates, which results in expensive computation when the number of clients is large. 
	
\subsubsection{Other model aggregation methods}
\label{subsubsec4.2.2}
Many security strategies have been proposed to deal with malicious clients who conduct untargeted attacks in federated learning. In order to demonstrate the effectiveness of a solution, the proposer of a security solution considers making his/her solution as defensive as possible against the attacking solution. Thus, the defense can be effective in untargeted attacks as well as partially effective in targeted attacks. Other aggregation models refer to aggregation schemes besides statistics-based schemes (such as similarity-based, voting-based, and trust-based mechanisms), and these approaches are described below.
	
FLTrust \cite{92}. FLTrust assumes that a small validation dataset is owned by the server. The server evaluates both the previous parameters and the parameters updated by the client in each iteration. First, the parameters of client uploading are considered as a vector, and the server uses the ReLU cosine similarity to determine the similarity of the client’s uploaded parameters. Then, the server normalizes each model parameter for each client. FLTrust essentially projects each local model onto a hypersphere of the vector space where the server model is located. Finally, the server computes the weighted average of the normalized local model update parameters as the global model. FLTrust uses a combination of techniques and has been demonstrated to outperform the statistical-based approach in defensive capabilities. This approach maintains the accuracy of the model even when 60$\%$ of the clients are malicious.
	
DnC \cite{97}. In this method, the server randomly selects k client updates to calculate the gradient. Then, the server calculates the average gradient values of these selected client updates. Using these values, the server computes the projection of the singular feature vectors in the direction of the gradient features. Then, the server removes the vector with the largest projection value and adds the remaining vectors to the benign dataset. Finally, the server calculates the average of all benign data as the updated parameters. Experiments in \cite{97} prove that this scheme is more effective than the benchmark method.
	
Ensemble federated learning \cite{19}. First, the server randomly divides all clients into k groups. The server trains models on each group separately and obtains k aggregated models. Then, the server uses these models to predict the labels of the samples and records the results predicted by each model. Finally, the server takes the models with mostly the same prediction results as benign models and uses them to compute the global model. This approach assumes that, among all of the clients, the proportion of malicious clients should be small. With a total of 1,000 clients on the MNIST dataset, this approach is able to maintain a correct rate of more than 80$\%$ with no more than 40 malicious clients. The scheme has high portability of profiles, but its defense is weak.
	
GAA \cite{100}. The GAA deployed on the server receives the client’s update gradient and determines a credit score for each client. The total credit of the GAA is finite, and it will give higher credit to the better-performing clients. Next, the GAA calculates a weight value for each client based on the credit score. Finally, the server calculates the global model based on this weight value and each client's update parameters. The core of this scheme is that GAA has a validation dataset that can determine whether the parameters uploaded by clients are benign or poisoned. This scheme is able to successfully converge in the case of 90$\%$ of malicious clients. In addition, this scheme assumes that both the validation dataset and the client dataset are IID.
	
attestedFL \cite{101}. This solution uses a fusion of multiple methods to determine whether a client is malicious or benign through multiple comparisons. First, the server records the history of clients’ updates. The server compares every update from the same client and removes the model that appears not to be trained. Second, the server simultaneously calculates the cosine similarity of the same update from different clients as a second basis for removing abnormal clients. Finally, the server evaluates whether the parameters updated by a client improve the global model by using its validation dataset. If not, the server removes the update parameters from that client. This solution has a high success rate of defense, but its drawback is the high computational complexity.

\section{Future directions}\label{sec5}
There are still many challenges in conducting poisoning attacks. First, the mathematical rationales for different methods of attack are not clear yet. For example, in a label reversal attack, the number of malicious clients required for a successful attack is not deterministic; this number is related to the data distribution. There is no mathematical explanation for this phenomenon. Second, research on poisoning attacks in different frameworks is not well-developed. Specifically, attack scenarios can be explored in conjunction with emerging federated learning frameworks (such as graph federated learning~\cite{72}, One-Shot federated learning~\cite{73}, and personalized federated learning~\cite{74}). Moreover, attack does not just mean destruction, it also means worth. For example, backdoor attacks can be combined with watermarking techniques to achieve the protection of intellectual property.
	
Concerning defense, defense methods that can satisfy strong defenses but have low overhead are still being explored. First, the balance between the protection of privacy and the traceability of the attacker is a problem. In other words, it is a challenge to achieve the best of both worlds between privacy and security. A solution that can protect data privacy and resist poisoning attacks is in demand. Second, the current research on defense methods is mainly in the field of horizontal federated learning. Research on defense methods in vertical federated learning and transfer federated learning has not started. Besides, since the realistic scenario computational power is limited, the defense methods that require less computational power to achieve the same defense capability are pursued more often. For poisoning attacks, defense methods have more research difficulty since the training data and process are invisible. Future research is needed to design methods that can defend against both poisoning attacks and inference attacks.
	
\section{Conclusions}\label{sec6}
The acceleration of global digitization, the emergence of the increasingly large volume of training data, and stringent privacy laws make federated learning become one of the most interesting research directions in the machine learning literature. Due to the invisibility of client-side training data and the Non-IID nature of these data, poisoning attacks on federated learning seem to be unavoidable. In this paper, we review how to conduct poisoning attacks and how to defend against them. More importantly, we summarize these poisoning attacks into different groups and evaluate the corresponding defenses. Finally, we point out the future directions of federated learning poisoning attacks and defenses.

\section*{Funding Statement}
This work was supported by Natural Science Foundations of Sichuan Province under Grants 2023NSFSC1397 and 2022NSFSC0552.

\bibliographystyle{elsarticle-num}

\bibliography{bibsample}

\begin{thebibliography}{10}
\expandafter\ifx\csname url\endcsname\relax
  \def\url#1{\texttt{#1}}\fi
\expandafter\ifx\csname urlprefix\endcsname\relax\def\urlprefix{URL }\fi
\expandafter\ifx\csname href\endcsname\relax
  \def\href#1#2{#2} \def\path#1{#1}\fi

\bibitem{1}
H.~B. McMahan, E.~Moore, D.~Ramage, B.~A. y~Arcas, Federated learning of deep
  networks using model averaging, arXiv preprint arXiv:1602.05629 2.

\bibitem{2}
K.~Bonawitz, H.~Eichner, W.~Grieskamp, D.~Huba, A.~Ingerman, V.~Ivanov,
  C.~Kiddon, J.~Kone{\v{c}}n{\`y}, S.~Mazzocchi, B.~McMahan, et~al., Towards
  federated learning at scale: System design, Proceedings of Machine Learning
  and Systems 1 (2019) 374--388.

\bibitem{3}
Q.~Li, Z.~Wen, Z.~Wu, S.~Hu, N.~Wang, Y.~Li, X.~Liu, B.~He, A survey on
  federated learning systems: Vision, hype and reality for data privacy and
  protection, IEEE Transactions on Knowledge and Data Engineering.

\bibitem{102}
J.~Kone{\v{c}}n{\`y}, B.~McMahan, D.~Ramage, Federated optimization:
  Distributed optimization beyond the datacenter, arXiv preprint
  arXiv:1511.03575.

\bibitem{4}
M.~Chen, A.~T. Suresh, R.~Mathews, A.~Wong, C.~Allauzen, F.~Beaufays, M.~Riley,
  Federated learning of n-gram language models, arXiv preprint
  arXiv:1910.03432.

\bibitem{5}
S.~Ramaswamy, R.~Mathews, K.~Rao, F.~Beaufays, Federated learning for emoji
  prediction in a mobile keyboard, arXiv preprint arXiv:1906.04329.

\bibitem{6}
T.~Yang, G.~Andrew, H.~Eichner, H.~Sun, W.~Li, N.~Kong, D.~Ramage, F.~Beaufays,
  Applied federated learning: Improving google keyboard query suggestions,
  arXiv preprint arXiv:1812.02903.

\bibitem{7}
M.~Liu, S.~Ho, M.~Wang, L.~Gao, Y.~Jin, H.~Zhang, Federated learning meets
  natural language processing: A survey, arXiv preprint arXiv:2107.12603.

\bibitem{8}
T.~S. Brisimi, R.~Chen, T.~Mela, A.~Olshevsky, I.~C. Paschalidis, W.~Shi,
  Federated learning of predictive models from federated electronic health
  records, International journal of medical informatics 112 (2018) 59--67.

\bibitem{9}
N.~Rieke, J.~Hancox, W.~Li, F.~Milletari, H.~R. Roth, S.~Albarqouni, S.~Bakas,
  M.~N. Galtier, B.~A. Landman, K.~Maier-Hein, et~al., The future of digital
  health with federated learning, NPJ digital medicine 3~(1) (2020) 1--7.

\bibitem{10}
A.~Saeed, T.~Ozcelebi, J.~Lukkien, Multi-task self-supervised learning for
  human activity detection, Proceedings of the ACM on Interactive, Mobile,
  Wearable and Ubiquitous Technologies 3~(2) (2019) 1--30.

\bibitem{11}
P.~Schmidt, A.~Reiss, R.~Duerichen, C.~Marberger, K.~Van~Laerhoven, Introducing
  wesad, a multimodal dataset for wearable stress and affect detection, in:
  Proceedings of the 20th ACM international conference on multimodal
  interaction, 2018, pp. 400--408.

\bibitem{12}
S.~Silva, B.~A. Gutman, E.~Romero, P.~M. Thompson, A.~Altmann, M.~Lorenzi,
  Federated learning in distributed medical databases: Meta-analysis of
  large-scale subcortical brain data, in: 2019 IEEE 16th international
  symposium on biomedical imaging, IEEE, 2019, pp. 270--274.

\bibitem{13}
A.~Supratak, H.~Dong, C.~Wu, Y.~Guo, Deepsleepnet: A model for automatic sleep
  stage scoring based on raw single-channel eeg, IEEE Transactions on Neural
  Systems and Rehabilitation Engineering 25~(11) (2017) 1998--2008.

\bibitem{14}
J.~Xu, B.~S. Glicksberg, C.~Su, P.~Walker, J.~Bian, F.~Wang, Federated learning
  for healthcare informatics, Journal of Healthcare Informatics Research 5~(1)
  (2021) 1--19.

\bibitem{64}
Q.~Yang, J.~Zhang, W.~Hao, G.~P. Spell, L.~Carin, Flop: Federated learning on
  medical datasets using partial networks, in: Proceedings of the 27th ACM
  SIGKDD Conference on Knowledge Discovery \& Data Mining, 2021, pp.
  3845--3853.

\bibitem{15}
P.~C.~M. Arachchige, P.~Bertok, I.~Khalil, D.~Liu, S.~Camtepe, M.~Atiquzzaman,
  A trustworthy privacy preserving framework for machine learning in industrial
  iot systems, IEEE Transactions on Industrial Informatics 16~(9) (2020)
  6092--6102.

\bibitem{16}
J.~Park, S.~Samarakoon, M.~Bennis, M.~Debbah, Wireless network intelligence at
  the edge, arXiv preprint arXiv:1812.02858.

\bibitem{17}
S.~Liu, T.~Li, Q.~Zhu, Communication-efficient distributed machine learning
  over strategic networks: A two-layer game approach, arXiv preprint
  arXiv:2011.01455.

\bibitem{18}
A.~N. Bhagoji, S.~Chakraborty, P.~Mittal, S.~Calo, Analyzing federated learning
  through an adversarial lens, in: International Conference on Machine
  Learning, PMLR, 2019, pp. 634--643.

\bibitem{19}
M.~Fang, X.~Cao, J.~Jia, N.~Gong, Local model poisoning attacks to
  byzantine-robust federated learning, in: 29th USENIX Security Symposium
  (USENIX Security 20), 2020, pp. 1605--1622.

\bibitem{20}
P.~Kairouz, H.~B. McMahan, B.~Avent, A.~Bellet, M.~Bennis, A.~N. Bhagoji,
  K.~Bonawitz, Z.~Charles, G.~Cormode, R.~Cummings, et~al., Advances and open
  problems in federated learning, Foundations and Trends{\textregistered} in
  Machine Learning 14~(1--2) (2021) 1--210.

\bibitem{21}
V.~Mothukuri, R.~M. Parizi, S.~Pouriyeh, Y.~Huang, A.~Dehghantanha,
  G.~Srivastava, A survey on security and privacy of federated learning, Future
  Generation Computer Systems 115 (2021) 619--640.

\bibitem{22}
Y.~Jin, X.~Wei, Y.~Liu, Q.~Yang, A survey towards federated semi-supervised
  learning, arXiv preprint arXiv:2002.11545.

\bibitem{23}
A.~Blanco-Justicia, J.~Domingo-Ferrer, S.~Mart{\'\i}nez, D.~S{\'a}nchez,
  A.~Flanagan, K.~E. Tan, Achieving security and privacy in federated learning
  systems: Survey, research challenges and future directions, Engineering
  Applications of Artificial Intelligence 106 (2021) 104468.

\bibitem{24}
Z.~Wang, Q.~Hu, Blockchain-based federated learning: A comprehensive survey,
  arXiv preprint arXiv:2110.02182.

\bibitem{25}
X.~Yin, Y.~Zhu, J.~Hu, A comprehensive survey of privacy-preserving federated
  learning: A taxonomy, review, and future directions, ACM Computing Surveys
  54~(6) (2021) 1--36.

\bibitem{26}
M.~Yang, Y.~He, J.~Qiao, Federated learning-based privacy-preserving and
  security: Survey, in: 2021 Computing, Communications and IoT Applications,
  IEEE, 2021, pp. 312--317.

\bibitem{27}
Z.~Wang, Q.~Kang, X.~Zhang, Q.~Hu, Defense strategies toward model poisoning
  attacks in federated learning: A survey, in: 2022 IEEE Wireless
  Communications and Networking Conference, IEEE, 2022, pp. 548--553.

\bibitem{28}
L.~Lyu, H.~Yu, X.~Ma, L.~Sun, J.~Zhao, Q.~Yang, P.~S. Yu, Privacy and
  robustness in federated learning: Attacks and defenses, arXiv preprint
  arXiv:2012.06337.

\bibitem{29}
Z.~Liu, J.~Guo, W.~Yang, J.~Fan, K.-Y. Lam, J.~Zhao, Privacy-preserving
  aggregation in federated learning: A survey, arXiv preprint arXiv:2203.17005.

\bibitem{30}
N.~Rodr{\'\i}guez-Barroso, D.~J. L{\'o}pez, M.~Luz{\'o}n, F.~Herrera,
  E.~Mart{\'\i}nez-C{\'a}mara, Survey on federated learning threats: Concepts,
  taxonomy on attacks and defences, experimental study and challenges, arXiv
  preprint arXiv:2201.08135.

\bibitem{32}
V.~Smith, C.-K. Chiang, M.~Sanjabi, A.~S. Talwalkar, Federated multi-task
  learning, Advances in neural information processing systems 30.

\bibitem{33}
Y.~Wu, S.~Cai, X.~Xiao, G.~Chen, B.~C. Ooi, Privacy preserving vertical
  federated learning for tree-based models, arXiv preprint arXiv:2008.06170.

\bibitem{65}
S.~Yang, B.~Ren, X.~Zhou, L.~Liu, Parallel distributed logistic regression for
  vertical federated learning without third-party coordinator, arXiv preprint
  arXiv:1911.09824.

\bibitem{34}
Y.~Chen, X.~Qin, J.~Wang, C.~Yu, W.~Gao, Fedhealth: A federated transfer
  learning framework for wearable healthcare, IEEE Intelligent Systems 35~(4)
  (2020) 83--93.

\bibitem{35}
E.~Bagdasaryan, V.~Shmatikov, Blind backdoors in deep learning models, in: 30th
  USENIX Security Symposium, 2021, pp. 1505--1521.

\bibitem{36}
E.~Bagdasaryan, A.~Veit, Y.~Hua, D.~Estrin, V.~Shmatikov, How to backdoor
  federated learning, in: International Conference on Artificial Intelligence
  and Statistics, PMLR, 2020, pp. 2938--2948.

\bibitem{37}
A.~N. Bhagoji, S.~Chakraborty, P.~Mittal, S.~Calo, Analyzing federated learning
  through an adversarial lens, in: International Conference on Machine
  Learning, PMLR, 2019, pp. 634--643.

\bibitem{38}
M.~Jagielski, G.~Severi, N.~Pousette~Harger, A.~Oprea, Subpopulation data
  poisoning attacks, in: Proceedings of the 2021 ACM SIGSAC Conference on
  Computer and Communications Security, 2021, pp. 3104--3122.

\bibitem{39}
Z.~Sun, P.~Kairouz, A.~Suresh, H.~McMahan, Can you really backdoor federated
  learning?, arXiv preprint arXiv:1911.07963.

\bibitem{40}
H.~Wang, K.~Sreenivasan, S.~Rajput, H.~Vishwakarma, S.~Agarwal, J.-y. Sohn,
  K.~Lee, D.~Papailiopoulos, Attack of the tails: Yes, you really can backdoor
  federated learning, Advances in Neural Information Processing Systems 33
  (2020) 16070--16084.

\bibitem{41}
A.~Du, B.~Chen, T.-J. Chin, Y.~W. Law, M.~Sasdelli, R.~Rajasegaran,
  D.~Campbell, Physical adversarial attacks on an aerial imagery object
  detector, in: Proceedings of the IEEE/CVF Winter Conference on Applications
  of Computer Vision, 2022, pp. 1796--1806.

\bibitem{42}
D.~Song, K.~Eykholt, I.~Evtimov, E.~Fernandes, B.~Li, A.~Rahmati, F.~Tramer,
  A.~Prakash, T.~Kohno, Physical adversarial examples for object detectors, in:
  12th USENIX workshop on offensive technologies, 2018.

\bibitem{43}
X.~Liu, H.~Yang, Z.~Liu, L.~Song, H.~Li, Y.~Chen, Dpatch: An adversarial patch
  attack on object detectors, arXiv preprint arXiv:1806.02299.

\bibitem{44}
J.~Zhang, B.~Chen, X.~Cheng, H.~T.~T. Binh, S.~Yu, Poisongan: Generative
  poisoning attacks against federated learning in edge computing systems, IEEE
  Internet of Things Journal 8~(5) (2020) 3310--3322.

\bibitem{49}
J.~Zhang, J.~Chen, D.~Wu, B.~Chen, S.~Yu, Poisoning attack in federated
  learning using generative adversarial nets, in: 2019 18th IEEE International
  Conference On Trust, Security And Privacy In Computing And
  Communications/13th IEEE International Conference On Big Data Science And
  Engineering, IEEE, 2019, pp. 374--380.

\bibitem{71}
J.~Gui, Z.~Sun, Y.~Wen, D.~Tao, J.~Ye, A review on generative adversarial
  networks: Algorithms, theory, and applications, IEEE Transactions on
  Knowledge and Data Engineering.

\bibitem{72}
K.~Zhang, C.~Yang, X.~Li, L.~Sun, S.~M. Yiu, Subgraph federated learning with
  missing neighbor generation, Advances in Neural Information Processing
  Systems 34 (2021) 6671--6682.

\bibitem{73}
Q.~Li, B.~He, D.~Song, Practical one-shot federated learning for cross-silo
  setting, arXiv preprint arXiv:2010.01017.

\bibitem{74}
A.~Shamsian, A.~Navon, E.~Fetaya, G.~Chechik, Personalized federated learning
  using hypernetworks, in: International Conference on Machine Learning, PMLR,
  2021, pp. 9489--9502.

\bibitem{51}
H.~Xiao, H.~Xiao, C.~Eckert, Adversarial label flips attack on support vector
  machines, Ecai 34~(34) (2014) 900--907.

\bibitem{52}
H.~Xiao, B.~Biggio, B.~Nelson, H.~Xiao, C.~Eckert, F.~Roli, Support vector
  machines under adversarial label contamination, Neurocomputing 160 (2015)
  53--62.

\bibitem{53}
V.~Tolpegin, S.~Truex, M.~E. Gursoy, L.~Liu, Data poisoning attacks against
  federated learning systems, in: European Symposium on Research in Computer
  Security, Springer, 2020, pp. 480--501.

\bibitem{55}
S.~Silva, B.~A. Gutman, E.~Romero, P.~M. Thompson, A.~Altmann, M.~Lorenzi,
  Federated learning in distributed medical databases: Meta-analysis of
  large-scale subcortical brain data, in: 2019 IEEE 16th international
  symposium on biomedical imaging, IEEE, 2019, pp. 270--274.

\bibitem{75}
N.~Carlini, Poisoning the unlabeled dataset of semi-supervised learning, in:
  30th USENIX Security Symposium, 2021, pp. 1577--1592.

\bibitem{76}
G.~Baruch, M.~Baruch, Y.~Goldberg, A little is enough: Circumventing defenses
  for distributed learning, Advances in Neural Information Processing Systems
  32.

\bibitem{77}
C.~Xie, O.~Koyejo, I.~Gupta, Fall of empires: Breaking byzantine-tolerant sgd
  by inner product manipulation, in: Uncertainty in Artificial Intelligence,
  PMLR, 2020, pp. 261--270.

\bibitem{78}
J.~Lin, M.~Du, J.~Liu, Free-riders in federated learning: Attacks and defenses,
  arXiv preprint arXiv:1911.12560.

\bibitem{79}
G.~Costa, F.~Pinelli, S.~Soderi, G.~Tolomei, Covert channel attack to federated
  learning systems, arXiv preprint arXiv:2104.10561.

\bibitem{103}
A.~Shafahi, W.~R. Huang, M.~Najibi, O.~Suciu, C.~Studer, T.~Dumitras,
  T.~Goldstein, Poison frogs! targeted clean-label poisoning attacks on neural
  networks, Advances in neural information processing systems 31.

\bibitem{48}
C.~Xie, K.~Huang, P.-Y. Chen, B.~Li, Dba: Distributed backdoor attacks against
  federated learning, in: International Conference on Learning Representations,
  2019.

\bibitem{50}
C.~Fung, C.~J.~M. Yoon, I.~Beschastnikh, Mitigating sybils in federated
  learning poisoning, arXiv preprint arXiv:1808.04866.

\bibitem{105}
B.~McMahan, E.~Moore, D.~Ramage, S.~Hampson, B.~A. y~Arcas,
  Communication-efficient learning of deep networks from decentralized data,
  in: Artificial intelligence and statistics, PMLR, 2017, pp. 1273--1282.

\bibitem{80}
C.~Dwork, A.~Roth, et~al., The algorithmic foundations of differential privacy,
  Foundations and Trends{\textregistered} in Theoretical Computer Science
  9~(3--4) (2014) 211--407.

\bibitem{60}
K.~Wei, J.~Li, M.~Ding, C.~Ma, H.~H. Yang, F.~Farokhi, S.~Jin, T.~Q. Quek,
  H.~V. Poor, Federated learning with differential privacy: Algorithms and
  performance analysis, IEEE Transactions on Information Forensics and Security
  15 (2020) 3454--3469.

\bibitem{81}
R.~C. Geyer, T.~Klein, M.~Nabi, Differentially private federated learning: A
  client level perspective, arXiv preprint arXiv:1712.07557.

\bibitem{dwork2014the}
C.~Dwork, A.~Roth, The algorithmic foundations of differential privacy,
  Foundations and Trends in Theoretical Computer Science 9 (2014) 211--407.

\bibitem{82}
M.~Du, R.~Jia, D.~Song, Robust anomaly detection and backdoor attack detection
  via differential privacy, arXiv preprint arXiv:1911.07116.

\bibitem{83}
B.~G. Doan, E.~Abbasnejad, D.~C. Ranasinghe, Februus: Input purification
  defense against trojan attacks on deep neural network systems, in: Annual
  Computer Security Applications Conference, 2020, pp. 897--912.

\bibitem{84}
T.~D. Nguyen, P.~Rieger, H.~Chen, H.~Yalame, H.~M{\"o}llering, H.~Fereidooni,
  S.~Marchal, M.~Miettinen, A.~Mirhoseini, S.~Zeitouni, et~al., Flame: Taming
  backdoors in federated learning, Cryptology ePrint Archive.

\bibitem{85}
L.~McInnes, J.~Healy, S.~Astels, hdbscan: Hierarchical density based
  clustering., J. Open Source Softw. 2~(11) (2017) 205.

\bibitem{86}
S.~Shen, S.~Tople, P.~Saxena, Auror: Defending against poisoning attacks in
  collaborative deep learning systems, in: Proceedings of the 32nd Annual
  Conference on Computer Security Applications, 2016, pp. 508--519.

\bibitem{57}
J.~Steinhardt, P.~W.~W. Koh, P.~S. Liang, Certified defenses for data poisoning
  attacks, Advances in neural information processing systems 30.

\bibitem{59}
B.~Wang, Y.~Yao, S.~Shan, H.~Li, B.~Viswanath, H.~Zheng, B.~Y. Zhao, Neural
  cleanse: Identifying and mitigating backdoor attacks in neural networks, in:
  2019 IEEE Symposium on Security and Privacy, IEEE, 2019, pp. 707--723.

\bibitem{58}
B.~Tran, J.~Li, A.~Madry, Spectral signatures in backdoor attacks, Advances in
  neural information processing systems 31.

\bibitem{88}
Y.~Khazbak, T.~Tan, G.~Cao, Mlguard: Mitigating poisoning attacks in privacy
  preserving distributed collaborative learning, in: 2020 29th International
  Conference on Computer Communications and Networks, IEEE, 2020, pp. 1--9.

\bibitem{89}
X.~Li, Z.~Qu, S.~Zhao, B.~Tang, Z.~Lu, Y.~Liu, Lomar: A local defense against
  poisoning attack on federated learning, IEEE Transactions on Dependable and
  Secure Computing.

\bibitem{54}
Y.~Shen, S.~Sanghavi, Learning with bad training data via iterative trimmed
  loss minimization, in: International Conference on Machine Learning, PMLR,
  2019, pp. 5739--5748.

\bibitem{90}
S.~Andreina, G.~A. Marson, H.~M{\"o}llering, G.~Karame, Baffle: Backdoor
  detection via feedback-based federated learning, in: 2021 IEEE 41st
  International Conference on Distributed Computing Systems, IEEE, 2021, pp.
  852--863.

\bibitem{91}
M.~S. Ozdayi, M.~Kantarcioglu, Y.~R. Gel, Defending against backdoors in
  federated learning with robust learning rate, in: Proceedings of the AAAI
  Conference on Artificial Intelligence, Vol.~35, 2021, pp. 9268--9276.

\bibitem{92}
X.~Cao, M.~Fang, J.~Liu, N.~Z. Gong, Fltrust: Byzantine-robust federated
  learning via trust bootstrapping, arXiv preprint arXiv:2012.13995.

\bibitem{62}
C.~Wu, X.~Yang, S.~Zhu, P.~Mitra, Mitigating backdoor attacks in federated
  learning, arXiv preprint arXiv:2011.01767.

\bibitem{87}
S.~Li, Y.~Cheng, W.~Wang, Y.~Liu, T.~Chen, Learning to detect malicious clients
  for robust federated learning, arXiv preprint arXiv:2002.00211.

\bibitem{106}
M.~Ester, H.-P. Kriegel, J.~Sander, X.~Xu, et~al., A density-based algorithm
  for discovering clusters in large spatial databases with noise., in: kdd,
  Vol.~96, 1996, pp. 226--231.

\bibitem{104}
L.~Lamport, R.~Shostak, M.~Pease, The byzantine generals problem, in:
  Concurrency: the works of leslie lamport, 2019, pp. 203--226.

\bibitem{93}
S.~Hong, V.~Chandrasekaran, Y.~Kaya, T.~Dumitra{\c{s}}, N.~Papernot, On the
  effectiveness of mitigating data poisoning attacks with gradient shaping,
  arXiv preprint arXiv:2002.11497.

\bibitem{94}
A.~Hard, K.~Rao, R.~Mathews, S.~Ramaswamy, F.~Beaufays, S.~Augenstein,
  H.~Eichner, C.~Kiddon, D.~Ramage, Federated learning for mobile keyboard
  prediction, arXiv preprint arXiv:1811.03604.

\bibitem{95}
P.~Blanchard, E.~M. El~Mhamdi, R.~Guerraoui, J.~Stainer, Machine learning with
  adversaries: Byzantine tolerant gradient descent, Advances in Neural
  Information Processing Systems 30.

\bibitem{97}
V.~Shejwalkar, A.~Houmansadr, Manipulating the byzantine: Optimizing model
  poisoning attacks and defenses for federated learning, in: Network and
  Distributed System Security Symposium, 2021.

\bibitem{96}
D.~Yin, Y.~Chen, R.~Kannan, P.~Bartlett, Byzantine-robust distributed learning:
  Towards optimal statistical rates, in: International Conference on Machine
  Learning, PMLR, 2018, pp. 5650--5659.

\bibitem{98}
M.~El~El~Mhamdi, R.~Guerraoui, S.~Rouault, The hidden vulnerability of
  distributed learning in byzantium, arXiv preprint arXiv:1802.07927.

\bibitem{100}
X.~Pan, M.~Zhang, D.~Wu, Q.~Xiao, S.~Ji, Z.~Yang, Justinian's gaavernor: Robust
  distributed learning with gradient aggregation agent, in: 29th USENIX
  Security Symposium, 2020, pp. 1641--1658.

\bibitem{101}
R.~A. Mallah, D.~Lopez, B.~Farooq, Untargeted poisoning attack detection in
  federated learning via behavior attestation, arXiv preprint arXiv:2101.10904.

\bibitem{99}
K.~Pillutla, S.~M. Kakade, Z.~Harchaoui, Robust aggregation for federated
  learning, arXiv preprint arXiv:1912.13445.

\end{thebibliography}

\end{document}